\documentclass[preprint]{aastex631} %

\usepackage{amsmath}
\DeclareMathOperator\erf{erf}

\submitjournal{Astrophysical Journal Letters}

\begin{document}

\title{Solar Flare Ion Temperatures}

\author[0000-0001-5690-2351]{Alexander J. B. Russell}
\affiliation{School of Mathematics \& Statistics, 
University of St Andrews, 
St Andrews, KY16 9SS, UK}

\correspondingauthor{Alexander J. B. Russell}
\email{ar51@st-andrews.ac.uk}

\author[0000-0002-4980-7126]{Vanessa Polito}
\affiliation{Lockheed Martin Solar \& Astrophysics Laboratory, Palo Alto, CA 94304, USA}
\affiliation{Department of Physics, Oregon State University, 301 Weniger Hall, Corvallis, OR 97331, USA}

\author[0000-0002-0405-0668]{Paola Testa}
\affiliation{Harvard-Smithsonian Center for Astrophysics, 60 Garden St, Cambridge, MA 02193, USA}

\author[0000-0002-8370-952X]{Bart De Pontieu}
\affiliation{Lockheed Martin Solar \& Astrophysics Laboratory, Palo Alto, CA 94304, USA}
\affiliation{Institute of Theoretical Astrophysics, University of Oslo, PO Box 1029, Blindern 0315, Oslo, Norway}
\affiliation{Rosseland Centre for Solar Physics, University of Oslo, PO Box 1029, Blindern 0315, Oslo, Norway}

\author[0000-0002-3505-9542]{Sergey A. Belov}
\affiliation{Centre for Fusion, Space and Astrophysics, Department of Physics, University of Warwick, Coventry, CV4 7AL, UK}

\begin{abstract}
This paper proposes that the ion temperature is several times the local electron temperature in the hot onset phase and at the above-the-loop region of solar flares. The paper considers: the evidence of spectral line Doppler widths (``non-thermal'' broadening); evidence for ``universal'' ion and electron temperature increase scaling relations for magnetic reconnection in the solar wind, Earth's magnetopause, Earth's magnetotail and numerical simulations; and thermal equilibration times for onset and above-the-loop densities, which are much longer than previous estimates based on soft X-ray flare loops. We conclude that the ion temperature is likely to reach 60~MK or greater and that it may represent a substantial part of spectral line widths, significantly contributing to solving the long-standing issue of the excess nonthermal broadening in flare lines.
\end{abstract}


\section{Introduction}\label{sec:intro}

The electron temperature $T_e$ in the solar corona can be determined by a variety of methods, including from emission line ratios and thermal X-ray emission. The inferred $T_e$ for solar flares typically ranges from 10~MK to 40~MK, which is consistent with the formation temperatures of highly-ionized ions found in flare spectra (the ionization fractions depend on $T_e$, not $T_i$).

The ion temperature $T_i$ is constrained by the Doppler temperature of spectral lines,
\begin{equation}
    T_D = \frac{m_i c^2}{2 k_B \lambda_0^2}\frac{\Delta \lambda_{fit}^2 - \Delta \lambda_{inst}^2}{4 \ln 2 },
\end{equation}
where $\Delta \lambda_{fit}$ is the observed FWHM width, $\Delta \lambda_{inst}$ quantifies the instrumental broadening and $m_i$ is the mass of the emitting ion. However, $T_D$ does not uniquely determine $T_i$ because it also contains contributions from unresolved line-of-sight mass motions. Thus, a $T_D$ measurement defines a parabola in a $(\xi,T_i)$ parameter space
\begin{equation}
    T_i = T_D - \frac{m_i \xi^2}{2k_B},
\end{equation}
where $\xi$ is the true non-thermal velocity.
This curve yields bounds $T_i \leq T_D$ and $\xi \leq v_{nt}$
where
\begin{equation}
    v_{nt} = \sqrt{\frac{2k_B (T_D-T_e)}{m_i}}
\end{equation}
is defined by assuming $T_i= T_e$. However, without further information, no point along the $(\xi,T_i)$ curve within these boundaries is more likely than any other.

\begin{figure}
\plotone{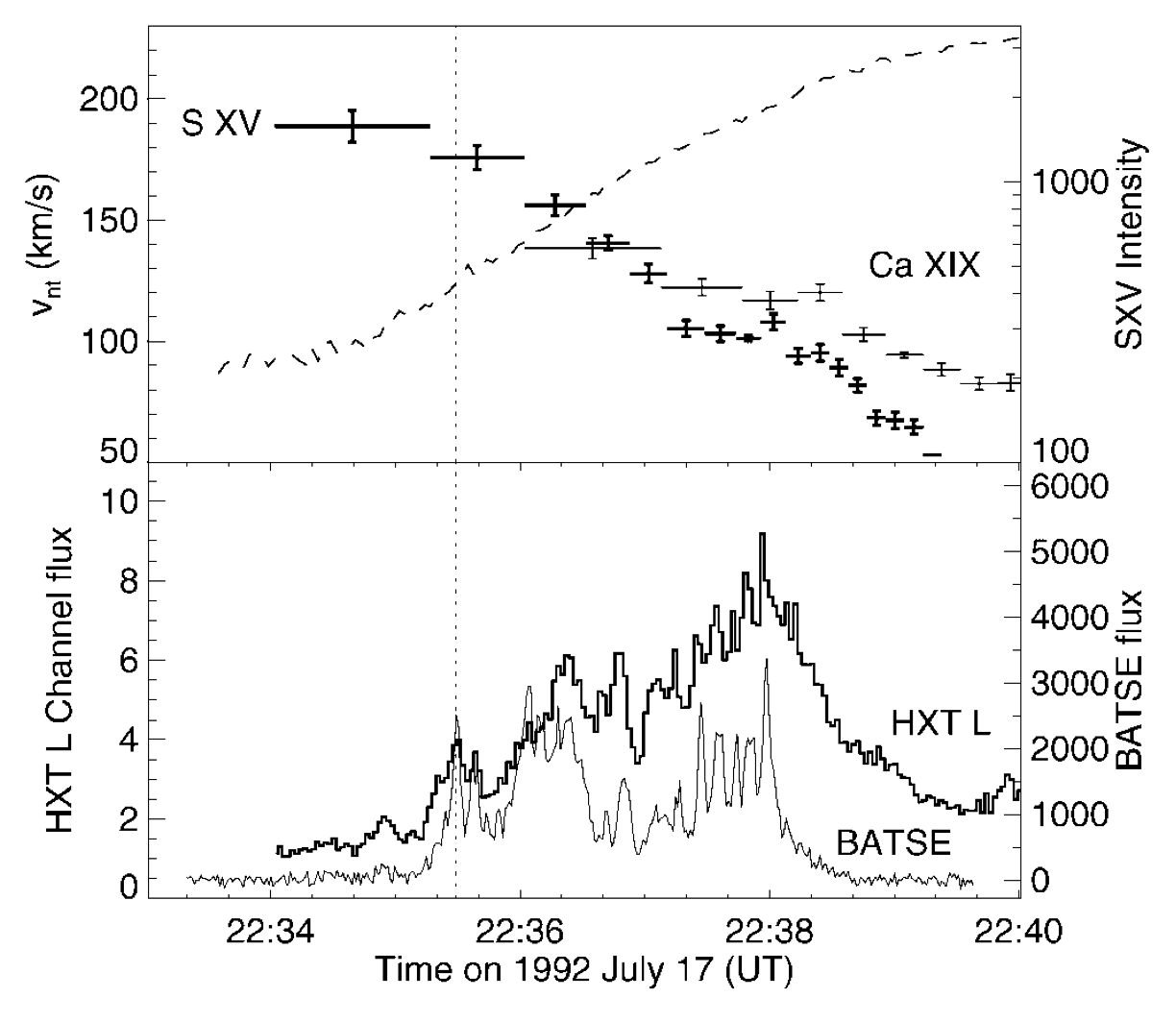}
\caption{Line broadening for the 1992 July 17 C5.3 flare, reproduced from \citet{1998Alexander}. (top) excess line width reported as nonthermal velocity for \ion{S}{15} and \ion{Ca}{19}. Horizontal error bars show the accumulation time to obtain each spectrum. The dashed line plots the \ion{S}{15} intensity, showing the rise of soft X-rays. The electron temperature (not plotted but used to derive $v_{nt}$) was determined using line ratio techniques.
(lower) hard X-rays showing the HXT L channel and BATSE (25–50 keV) fluxes. Reproduced by permission of the AAS.}\label{fig:alexander}
\end{figure}

Flare lines of ions with formation temperatures above $10\mbox{ MK}$ are much broader than would be expected if $T_i$ were equal to $T_e$. This phenomenon was discovered in whole-Sun high-spectral-resolution X-ray spectra by \textit{P78-1} \citep{1979Doschek}, \textit{SMM} \citep{1981Culhane} and \textit{Hinotori} \citep{1982Tanaka} and later studied with improved instrumentation by \textit{Yohkoh} BCS \citep{1991Culhane}. For reviews, see \citet{1990Doschek} and \citet{1999Antonucci}. The X-ray line profiles are broadest before the impulsive phase \citep{1979Doschek}, when ion Doppler temperatures can be $\gtrsim 60$~MK. For comparison, during the hot onset phase $T_e$ is typically $10$--$15\mbox{ MK}$ \citep{2021Hudson,2023Battaglia}. In the impulsive and gradual phases, $T_D$ decreases towards $T_e$, as shown in Figure~\ref{fig:alexander}. For other examples, see Figure~4 of \citet{1986Tanaka} or Figure~1 of \citet{1998Alexander}.

\textit{Hinode} EIS \citep{2007Culhane} and \textit{IRIS} \citep{2014DePontieu,2021DePontieu} have studied extreme ultraviolet (EUV) line widths with spatial resolution. The broadening occurs in a wide range of conditions, which is difficult to reconcile with pressure or opacity broadening. Reviews can be found in \citet{2024Russell} and \citet{2024Fletcher}. EIS maps of coronal line widths during the main stages of the flare have found that line widths are greater at the above-the-loop region ($v_{nt}\gtrsim100\mbox{ km s}^{-1}$) than in the underlying flare loops \citep{2008Hara,2013Harra,2014Doschek,2017Kontar,2021Stores}. Line widths in the above-the-loop region have also been studied by \textit{IRIS}, e.g., \citet{2023Shen} and \citet{2024Ashfield}.

Several early flare line studies referred directly to ion temperature, e.g. ``Alternatively the line broadening may be interpreted as a decrease of ion temperature from $40\times10^6$ K to $20\times10^6$ K" \citep{1981Culhane}, or figures in \citet{1982Antonucci} and \citet{1986Tanaka}. However, in the 1980s a consensus developed of taking $T_i=T_e$, attributing the difference between $T_e$ and $T_D$ to unresolved mass motions and reporting the value of $v_{nt}$, e.g. discussion at the \textit{SMM} Flare Workshop recorded in \citet{1986Doschek}.

The belief that line broadening is due to unresolved motions has come under pressure from efforts to identify the nature of the motions responsible. Historical observations of soft X-ray lines were broad several minutes before evaporative upflows were detected, inconsistent with unresolved chromospheric evaporation \citep[][and Figure~1]{1983AntonucciDennis,1984Antonucci,1986Antonucci,1986Doschek}. Modeling of \textit{IRIS} \ion{Fe}{21} 1354.1~{\AA} lines, which are mostly symmetric, also indicates that line shapes are incompatible with explanation by unresolved upflows \citep{2019Polito}. More generally, motions in magnetized plasma typically imply that the broadening should depend on the viewing angle between the magnetic field direction and line of sight. Center-to-limb studies using uncollimated soft X-ray spectrometers found no significant viewing angle effect for widths of hot flare lines, despite a viewing angle effect existing for blueshifts due to chromospheric evaporation \citep{1984Antonucci,1986Antonucci,1986Doschek,1993Mariska,1994Mariska}. 

This paper critiques the $T_i = T_e$ assumption for the hot onset phase and in the above-the-loop region during the main flare. We propose that the data may be simply explained by applying ion and electron heating scaling laws from space plasma measurements and kinetic simulations, noting that the ion-electron thermal equilibration times are reasonably long for the relevant densities.

\newpage

\section{Empirical Heating Relations for Reconnection}\label{sec:ion-heating}

\subsection{In-situ Evidence from Space Plasmas}

In-situ measurements of reconnection exhausts across the solar wind, magnetopause and magnetotail display universal scaling laws for ion and electron heating. Analyzing 22 exhausts encountered by \textit{Wind} at Earth's L1 Lagrange point, \citet{2009Drake} found that the bulk temperature increase for protons follows
\begin{equation}\label{eq:Ti_empirical}
    \Delta T_p = 0.13 m_p V_A^2,
\end{equation}
where temperature is expressed in energy units, $m_p$ is the proton mass and $V_A$ is the upstream Alfvén speed based on the reconnecting magnetic field component. The factor $m_p V_A^2 = B^2/ \mu_0 n_e$ measures the free magnetic energy per electron-proton pair, so the 0.13 coefficient reflects the fraction of the available energy that goes into bulk proton heating. The same scaling was confirmed in larger solar wind studies using 188 \textit{Wind} events at L1 \citep{2017Mistry} and 88 \textit{Helios} events covering heliospheric distances from 0.3 to 1~au \citep{2020Tilquin}. A statistical survey of dayside magnetopause reconnection events by \citet{2014Phan} using \textit{THEMIS} found the same relation.

For electrons, \citet{2013Phan} found an empirical relation $\Delta T_e = 0.017 m_p V_A^2$ for magnetopause reconnection. Comparing the electron and proton heating, \citet{2014Phan} found $\Delta T_p / \Delta T_e \sim 8$, consistent with the value of 7.6 obtained from dividing the proton and electron temperature increase relations. 

The empirical data have now been extended to the high-Alfvén-speed low-$\beta$ regime relevant to solar flares by a statistical survey of reconnection events in Earth's magnetotail using \textit{MMS}. In this context, \citet{2023Oieroset} found that the electron bulk heating follows 
\begin{equation}\label{eq:Te_empirical}
\Delta T_e = 0.02 m_p V_A^2,
\end{equation} 
consistent with \citet{2013Phan} but with a slight increase to the coefficient and tested for upstream Alfvén speeds up to 4000~$\mbox{km s}^{-1}$. 
For ions, \citet{2024Oieroset} found that Eq~(\ref{eq:Ti_empirical})
holds well at lower upstream Alfvén speeds, yielding a ratio 
\begin{equation}
    \Delta T_p/ \Delta T_e \approx 6.5 \approx (m_e/m_p)^{1/4}.
\end{equation}
For an interesting discussion of this relation and its mass ratio interpretation, also addressing preferential ion heating by shocks, see \citet{2025Oka}. For the most strongly heated reconnection events, $\Delta T_p$ undershoots the linear scaling, reducing the ratio $\Delta T_p / \Delta T_e$. Nonetheless, $\Delta T_p / \Delta T_e$ remains above 4 for most of the higher-Alfvén-speed events analyzed by \citet{2024Oieroset}.

Equations~(\ref{eq:Ti_empirical}) and (\ref{eq:Te_empirical}) imply that 15\% of the available magnetic energy goes into heating the reconnection exhaust. The remaining 85\% is mainly converted to bulk kinetic energy of the reconnection outflow jets, as discussed by \citet{2019Zhang}.

\subsection{Mechanism and Simulations}

\citet{2009Drake} proposed that ions receive an initial Fermi kick when they first enter a reconnection exhaust, which boosts their thermal speed to a fraction of the reconnection outflow speed, in a process similar to heating of pickup ions. Since $m_e\ll m_p$, the energy that electrons gain on entering the reconnection exhaust is negligible compared to that gained by the ions. Hence, the initial energization of electrons is governed by less efficient processes within the reconnection layer.

\citet{2009Drake} verified the pickup Fermi mechanism for anti-parallel reconnection using test ions in 2D particle-in-cell (PIC) simulations. Later simulations by \citet{2009DrakeApJ}, \citet{2011Knizhnik} and \citet{2014DrakeSwisdak} found that ion heating is suppressed for reconnection with weak shear, which has also been seen in solar wind data analyzed by \citet{2020Tilquin}. The scaling of electron heating in 2.5D PIC simulations was investigated by \citet{2014Shay} and \citet{2015Haggerty}, who confirmed that ions are preferentially heated under a wide range of conditions. For 3D simulations, see \citet{2015Dahlin,2017Dahlin}, \citet{2021Zhang} and \citet{2024Yin}.

The current generation of 3D MHD flare simulations \citep[e.g.][]{2019Cheung,2023Ruan,2023Shen} typically solve single-temperature MHD and therefore do not treat preferential ion heating by magnetic reconnection. Capturing this effect in future MHD flare simulations will require development of multi-temperature models that successfully parameterize the small-scale physics involved in the ion and electron heating.

\subsection{Application to Solar Flares}

If the empirical temperature increase relations from space plasmas and kinetic simulations are applied to coronal reconnection, then setting a coronal value of $V_A = 2000~\mbox{km s}^{-1}$ in Eq~(\ref{eq:Te_empirical}) yields $\Delta T_e =9.6\mbox{ MK}$, which is consistent with an 11~MK hot onset temperature \citep{2021Hudson,2023Battaglia} after adding an initial temperature of 1--2~MK. Equation~(\ref{eq:Ti_empirical}) predicts a corresponding $\Delta T_p=62\mbox{ MK}$. The $\Delta T_p$ reduces to 38~MK if one applies a lower ratio $\Delta T_p / \Delta T_e = 4$, so we consider a range of approximately 40--65~MK for the expected $T_p$. 

The proton temperatures obtained are broadly consistent with the Doppler temperatures of flare lines without needing to invoke large non-thermal motions. If all ions are assumed to have the proton temperature, then the equivalent $v_{nt}$ ranges for this worked example are 93--127 $\mbox{ km s}^{-1}$ for iron, 110--150 $\mbox{ km s}^{-1}$ for calcium and 123--168 $\mbox{ km s}^{-1}$ for sulfur.
\citet{2009Drake} presented evidence that more massive ions are preferentially heated in solar wind reconnection exhausts. Thus, departures from thermal-equilibrium between the ion species would raise the estimated $v_{nt}$ ranges and potentially change the ordering by ion mass. This may explain why \ion{Ca}{19} often has a higher reported $v_{nt}$ than \ion{S}{15} \citep[e.g. Figure~1 of][]{1998Alexander}.

\section{Densities}\label{sec:densities}

The historic justification for assuming $T_i=T_e$ in flares was that flare time scales are much longer than the thermal equilibration time between species, assuming the electron density of soft X-ray plasma \citep{1984Antonucci,1984AntonucciB,1986Doschek}. A few years previously, \citet{1980McKenzie} and \citet{1981Doschek} measured the density of 2~MK flare plasma using the \ion{O}{7} $\lambda22.10$~{\AA}/$\lambda21.80$~{\AA} line ratio, reporting the peak density exceeded $10^{11}\mbox{ cm}^{-3}$ 1979 June~10 X2 flare, and that it exceeded $10^{12}\mbox{ cm}^{-3}$ for the 1980 April~8 M4 flare. A density of $10^{12}\mbox{ cm}^{-3}$ corresponds to an electron-proton thermalization time of approximately 1~s at flare temperatures, which is indeed short compared to impulsive phase durations. 

However, such high densities are only achieved in flare loops that have been filled by chromospheric evaporation \citep[e.g.][]{2011Fletcher}. In the standard flare model, densities are several orders of magnitude lower at the onset stage (before chromospheric evaporation is observed) and in the above-the-loop region where field lines have recently undergone reconnection but have not yet been filled with hot dense plasma of chromospheric origin.

For the above-the-loop region during the main stages of the flare, \citet{2008Hara} estimated a density of $6\times10^9\mbox{ cm}^{-3}$ using emission measure estimates. For a larger sample of events, \citet{1997Aschwanden} estimated that densities at the electron acceleration site are $6\times10^8$ to $10^{10}\mbox{ cm}^{-3}$ from radio observations. 

For the onset phase, we consider standard coronal densities of not more than $10^{10}\mbox{ cm}^{-3}$. \citet{1983Benz} reported broadband decimetric radiation before the impulsive phase from 400 to 1000~MHz, which corresponds to densities from $2\times10^9$ to $10^{10}\mbox{ cm}^{-3}$, within the range of values we consider.

\section{Thermal Equilibration}\label{sec:nu}

Thermal equilibration in a multi-species plasma is governed by
\begin{equation}\label{eq:dT}
    \frac{d T_\alpha}{dt} = \sum_{\beta\neq\alpha} \nu^\epsilon_{th,\alpha\beta}(T_\beta-T_\alpha),
\end{equation}
where the equilibration rates are 
\begin{equation}\label{eq:nu}
    \nu^\epsilon_{th,\alpha\beta} = 6.935\times10^{-21} \frac{(m_\alpha m_\beta)^{1/2}Z_\alpha^2 Z_\beta^2\lambda_{\alpha\beta}}{\left(m_\beta T_\alpha + m_\alpha T_\beta\right)^{3/2}} n_\beta.
\end{equation}
The above equations are given in SI units with temperatures in kelvin and particle charges $Z$ in units of $e$. Equivalent expressions are given in the NRL Plasma Formulary and \citet{2015Tracy}. Appendix \ref{sec:coulomb_log} provides formulas for the Coulomb logarithm $\lambda_{\alpha\beta}$, abundances used in our calculations and a remark on relative drifts.

Table~\ref{tab:nu:multi} calculates thermal equilibration rates for flare parameters, assuming $T_i =  60 \text{ MK}$ for all ions and $T_e =  15 \text{ MK}$. The stated rates are for $n_e = 10^{10} \text{ cm}^{-3}$ and should be scaled by multiplying by $n_e$ ($10^{10} \text{ cm}^{-3}$). \ion{S}{15}, \ion{Ca}{19} and \ion{Fe}{21} have been included as a sample of heavy ions. For columns where the field particles $\beta$ are heavy ions, the stated values should also be multiplied by the ionic fraction, e.g., $n_\text{Fe XXI}/n_\text{Fe}\leq1$. For simplicity, we consider the most common mass isotopes.

\begin{deluxetable}{c|cccccc}[h]
\tablecolumns{7}
\tablewidth{\columnwidth} 
\label{tab:nu:multi}
\tablecaption{$\nu^\epsilon_{\alpha\beta}$ for $T_e = 15 \text{ MK}$ and $T_i = 60 \text{ MK}$.} 
\tablehead{ Species  & \multicolumn{6}{c}{Field particles $\beta$} \\
$\alpha$ & e & p & \ion{He}{3} & \ion{S}{15} & \ion{Ca}{19} & \ion{Fe}{21}
}
\tablehead{ Species  & \multicolumn{6}{c}{Field particles $\beta$} \\
$\alpha$ & e & p & \ion{He}{3} & \ion{S}{15} & \ion{Ca}{19} & \ion{Fe}{21}
}
\startdata
    e            & --    & 0.011 & 1.3E-3 & 5.3E-7 & 5.1E-7 & 6.2E-6 \\
    p            & 0.013 & --    & 6.1E-3 & 3.0E-6 & 2.9E-6 & 3.6E-5 \\
    \ion{He}{3}  & 0.013 & 0.050 & --     & 2.1E-5 & 2.1E-5 & 2.6E-4 \\ 
    \ion{S}{15}  & 0.082 & 0.38  & 0.32   & --     & 1.4E-3 & 0.020  \\
    \ion{Ca}{19} & 0.11  & 0.50  & 0.43   & 1.9E-3 & --     & 0.032  \\
    \ion{Fe}{21} & 0.094 & 0.44  & 0.39   & 1.9E-3 & 2.3E-3 & --     \\
\enddata  
\tablecomments{Values should be multiplied by $n_e$ ($10^{10} \text{ cm}^{-3}$). Columns for heavy ions should additionally be multiplied by the corresponding ionic fraction, e.g., $n_\text{Fe XXI}/n_\text{Fe}$.}
\end{deluxetable}
\vspace{-1.5cm}

The first row of Table~\ref{tab:nu:multi} shows that electrons primarily thermalize with protons (column with the largest $\nu^\epsilon_{e\beta}$). The second row shows that protons primarily thermalize with electrons. The heavy ions are of negligible importance as fields, which can be attributed to their low densities. Simplifying Eq.~(\ref{eq:nu}) using $T_p/T_e \ll m_p/m_e$ shows that $\nu_{pe}^{\epsilon}\propto n_e T_e^{-3/2}$, hence the proton-electron thermalization time is insensitive to the proton temperature. 

Examining the other interactions, the temperatures of the heavy ions equilibrate to the proton temperature on a short time scale of several seconds, which suggests that a two-temperature model with a single temperature for all ions is a reasonable first approximation. Interestingly, \ion{He}{3} ions (alpha particles) equilibrate towards the proton temperature on an intermediate timescale, which may motivate three-temperature modeling.

Using the onset and above-the-loop densities of $6\times10^8$ to $10^{10}\mbox{ cm}^{-3}$ noted in \S\ref{sec:densities} yields electron-proton thermal equilibration time scales $(\nu_{ep}^{\epsilon})^{-1}\approx(\nu_{pe}^{\epsilon})^{-1}$ of approximately 130 to 1300~s, which implies that temperature differences between ions and electrons are not prevented and that once created they may persist for a long period in a given plasma element. These time scales do not place an upper limit on how long $T_i>T_e$ can occur in a flare since reconnection sequentially heats different plasma elements.

\section{Conclusion}\label{sec:conc}

This paper has proposed that magnetic reconnection in solar flares preferentially heats ions, producing $T_i>T_e$ in the hot onset stage and at the above-the-loop region. This proposition is based on the ``universal'' heating laws for reconnection that have recently been discovered in space plasmas and kinetic simulations, which predict $T_i/T_e$ is between 4 and 6.5 in reconnection exhausts in low-collisionality plasmas (\S\ref{sec:ion-heating}).

Temperature differences between electrons and ions can exist at onset and above-the-loop densities because of the relatively long thermal equilibration times, ranging from hundreds to thousands of seconds. The previous view that thermal equilibration times were short for flares was based on the high densities of soft X-ray loops that have been filled by chromospheric evaporation, which are approximately 10 to 200 times more dense than the onset and above-the-loop plasma (\S\ref{sec:densities} and \ref{sec:nu}).

This new framework offers a simple resolution to long-standing puzzles. Hot onsets exhibit a steady 10--15~MK electron temperature with rising emission measure. The steady $T_e$ can be explained by the heating scaling relations if the Alfvén speed based on the reconnecting field component upstream of the reconnection layer remains fairly steady during the onset stage. The predicted $T_i$ would be 4 to 6.5 times $T_e$, which is consistent with observed ion Doppler temperatures. Potentially, this could account for most of the excess line width, although unresolved motions such as turbulence could also contribute. An explanation of ``excess'' line widths using the ion temperature also fits well with evidence that flare line widths do not exhibit center-to-limb variation and that the line cores are mostly symmetric. Departures from thermal-equilibrium between ion species combined with preferential heating of the most massive ions could potentially explain why \ion{Ca}{19} often has a higher reported $v_{nt}$ than \ion{S}{15}.

Once the flare begins, turbulence is expected to be produced in the above-the-loop region by braking of the reconnection outflow jet, and it is likely needed in the impulsive phase to accelerate particles in the above-the-loop region \citep{1993Larosa,1994Larosa,1995Tsuneta,2017Kontar,2021Stores,2023Ruan,2023Shen,2023Shibata,2024Russell,2024French}. However, the amplitude of this turbulence and the energy that goes into waves/Poynting flux will need to be revised downwards if the ions are hotter than the electrons. 

Our conclusion creates new demand for techniques that can separate line width contributions from $T_i$ and the true nonthermal velocity under flare conditions. Relevant techniques have been developed for the quiet Sun, quiescent active regions and coronal holes by \citet{1997Seely}, \citet{1998Tu}, \citet{2007Landi}, \citet{2008DollaSolomon}, \citet{2009Imada} and \citet{2023Zhu}, who found evidence of $T_i>T_e$ in those environments. Those studies provide a starting point for future technique development, although we remark that they assumed that all ions share the same true non-thermal velocity and we would be cautious about making the same assumption for flares due to very strong spatial gradients. 

Multi-temperature fluid models are widely used across different branches of plasma physics, including planetary ionospheres and the solar wind \citep[e.g.][]{2014vanderHolst,2022vanderHolst}, and our work motivates greater use of such models in solar physics. As an illustration of the potential importance beyond solar flares, \citet{2025Belov} recently found that the damping time for standing slow waves in hot coronal loops changes by 50\% when comparing one- and two-temperature models. For flare modeling specifically, there is now a need for multi-temperature fluid models that parameterize small-scale effects in such a way as to reproduce the ion and electron temperature increases seen in kinetic simulations and observations of magnetic reconnection.

Finally, we advocate for advancing mapping and reporting of ion Doppler temperatures and thermal equilibration rates to constrain where and when $T_i$ is likely to depart from $T_e$. The upcoming \textit{MUSE} \citep{2020DePontieu,2022Cheung} and \textit{Solar-C} EUVST \citep{2019Shimizu} missions are excellently suited to these tasks.

\vspace{5mm}

The authors thank the referee for valuable comments that improved the manuscript, and Sarah Matthews for author permission to reproduce Figure~1.
A.J.B.R. acknowledges support from STFC grant ST/W001195/1. V.P., P.T. and A.J.B.R. acknowledge NASA HGI grant No. 80NSSC20K0716. B.D.P., V.P. and P.T. were supported by NASA contract 80GSFC21C0011 (MUSE). P.T. was supported by contracts 4105785828 (MUSE), 8100002705 (IRIS), and NASA contract NNM07AB07C (Hinode/XRT) to the Smithsonian Astrophysical Observatory, and by NASA grant 80NSSC21K1684. S.A.B. acknowledges funding by STFC Grant ST/X000915/1. We thank the organizers of the 11th Coronal Loops Workshop (June 2024) and the 2025 US-Japan Workshop on Magnetic Reconnection (March 2025) for providing environments that facilitated this work. This research has made use of NASA's Astrophysics Data System Bibliographic Services.

\appendix

\section{Coulomb Logarithms}\label{sec:coulomb_log}

For ion-ion collisions the Coulomb logarithm is
\begin{equation}\label{eq:log_ii}
    \lambda_{\alpha,\beta} = 15.9 - \ln\left(\frac{Z_\alpha Z_\beta(m_\alpha+m_\beta)}{m_\beta T_\alpha + m_\alpha T_\beta}\sqrt{\frac{n_\alpha Z_\alpha^2}{T_\alpha}+\frac{n_\beta Z_\beta^2}{T_\beta}}\right),
\end{equation}
and for ion-electron collisions with $T_i < 2.1 \times 10^8 Z^2/\mu_i$,
\begin{equation}\label{eq:log_ei}
    \lambda_{\alpha,\beta} = \begin{cases} 
    21.5 - \ln\left(n_e^{1/2}T_e^{-1}\right), 
        & T_e > 1.2\times 10^5 Z^2 \\
    15.9 - \ln\left(n_e^{1/2} Z T_e^{-3/2} \right), 
        & T_e < 1.2\times 10^5 Z^2.
    \end{cases}.
\end{equation}

Elemental abundances used to calculate densities were curated from the flare literature, adopting A(He) = 11.09 \citep{2005Feldman}, A(S) = 6.90 \citep{1999FludraSchmelz}, A(Ca) = 6.77 \citep{2022Sylwester} and A(Fe) = 7.91 \citep{2012PhillipsDennis}, which are close to the photospheric abundances given by \citet{2009Asplund}. 

Equation~(\ref{eq:nu}) assumes that normalized relative drifts are negligible, $x=|\boldsymbol{U}_\alpha - \boldsymbol{U}_\beta|/\sqrt{v_{th,\alpha}^2 + v_{th,\beta}^2}\to0$. If relative drifts were significant then $\nu^\epsilon_{th,\alpha\beta}$ would be multiplied by a factor proportional to $\sqrt{\pi}\erf(x)/2x \leq 1$, which further helps to maintain temperature differences \citep{1985HernandezMarsch}. 

\bibliography{ThermalEquilibration}{}

\begin{thebibliography}{}
\expandafter\ifx\csname natexlab\endcsname\relax\def\natexlab#1{#1}\fi
\providecommand{\url}[1]{\href{#1}{#1}}
\providecommand{\dodoi}[1]{doi:~\href{http://doi.org/#1}{\nolinkurl{#1}}}
\providecommand{\doeprint}[1]{\href{http://ascl.net/#1}{\nolinkurl{http://ascl.net/#1}}}
\providecommand{\doarXiv}[1]{\href{https://arxiv.org/abs/#1}{\nolinkurl{https://arxiv.org/abs/#1}}}

\bibitem[{{Alexander} {et~al.}(1998){Alexander}, {Harra-Murnion}, {Khan}, \& {Matthews}}]{1998Alexander}
{Alexander}, D., {Harra-Murnion}, L.~K., {Khan}, J.~I., \& {Matthews}, S.~A. 1998, \apjl, 494, L235, \dodoi{10.1086/311175}

\bibitem[{{Antonucci}(1984)}]{1984AntonucciB}
{Antonucci}, E. 1984, \memsai, 55, 699

\bibitem[{{Antonucci} {et~al.}(1999){Antonucci}, {Alexander}, {Culhane}, {de Jager}, {MacNeice}, {Somov}, \& {Zarro}}]{1999Antonucci}
{Antonucci}, E., {Alexander}, D., {Culhane}, J.~L., {et~al.} 1999, in The many faces of the sun: a summary of the results from NASA's Solar Maximum Mission., ed. K.~T. {Strong}, J.~L.~R. {Saba}, B.~M. {Haisch}, \& J.~T. {Schmelz}, 331--391

\bibitem[{{Antonucci} \& {Dennis}(1983)}]{1983AntonucciDennis}
{Antonucci}, E., \& {Dennis}, B.~R. 1983, \solphys, 86, 67, \dodoi{10.1007/BF00157175}

\bibitem[{{Antonucci} {et~al.}(1984){Antonucci}, {Gabriel}, \& {Dennis}}]{1984Antonucci}
{Antonucci}, E., {Gabriel}, A.~H., \& {Dennis}, B.~R. 1984, \apj, 287, 917, \dodoi{10.1086/162749}

\bibitem[{{Antonucci} {et~al.}(1986){Antonucci}, {Rosner}, \& {Tsinganos}}]{1986Antonucci}
{Antonucci}, E., {Rosner}, R., \& {Tsinganos}, K. 1986, \apj, 301, 975, \dodoi{10.1086/163960}

\bibitem[{{Antonucci} {et~al.}(1982){Antonucci}, {Gabriel}, {Acton}, {Culhane}, {Doyle}, {Leibacher}, {Machado}, {Orwig}, \& {Rapley}}]{1982Antonucci}
{Antonucci}, E., {Gabriel}, A.~H., {Acton}, L.~W., {et~al.} 1982, \solphys, 78, 107, \dodoi{10.1007/BF00151147}

\bibitem[{{Aschwanden} \& {Benz}(1997)}]{1997Aschwanden}
{Aschwanden}, M.~J., \& {Benz}, A.~O. 1997, \apj, 480, 825, \dodoi{10.1086/303995}

\bibitem[{{Ashfield} {et~al.}(2024){Ashfield}, {Polito}, {Yu}, {Collier}, \& {Hayes}}]{2024Ashfield}
{Ashfield}, W., {Polito}, V., {Yu}, S., {Collier}, H., \& {Hayes}, L.~A. 2024, \apj, 973, 96, \dodoi{10.3847/1538-4357/ad64ca}

\bibitem[{{Asplund} {et~al.}(2009){Asplund}, {Grevesse}, {Sauval}, \& {Scott}}]{2009Asplund}
{Asplund}, M., {Grevesse}, N., {Sauval}, A.~J., \& {Scott}, P. 2009, \araa, 47, 481, \dodoi{10.1146/annurev.astro.46.060407.145222}

\bibitem[{{Battaglia} {et~al.}(2023){Battaglia}, {Hudson}, {Warmuth}, {Collier}, {Jeffrey}, {Caspi}, {Dickson}, {Saqri}, {Purkhart}, {Veronig}, {Harra}, \& {Krucker}}]{2023Battaglia}
{Battaglia}, A.~F., {Hudson}, H., {Warmuth}, A., {et~al.} 2023, \aap, 679, A139, \dodoi{10.1051/0004-6361/202347706}

\bibitem[{{Belov} {et~al.}(2025){Belov}, {Goffrey}, {Arber}, \& {Kolotkov}}]{2025Belov}
{Belov}, S.~A., {Goffrey}, T., {Arber}, T.~D., \& {Kolotkov}, D.~Y. 2025, \aap, 693, A186, \dodoi{10.1051/0004-6361/202452938}

\bibitem[{{Benz} {et~al.}(1983){Benz}, {Barrow}, {Dennis}, {Pick}, {Raoult}, \& {Simnett}}]{1983Benz}
{Benz}, A.~O., {Barrow}, C.~H., {Dennis}, B.~R., {et~al.} 1983, \solphys, 83, 267, \dodoi{10.1007/BF00148280}

\bibitem[{{Cheung} {et~al.}(2019){Cheung}, {Rempel}, {Chintzoglou}, {Chen}, {Testa}, {Mart{\'\i}nez-Sykora}, {Sainz Dalda}, {DeRosa}, {Malanushenko}, {Hansteen}, {De Pontieu}, {Carlsson}, {Gudiksen}, \& {McIntosh}}]{2019Cheung}
{Cheung}, M.~C.~M., {Rempel}, M., {Chintzoglou}, G., {et~al.} 2019, Nature Astronomy, 3, 160, \dodoi{10.1038/s41550-018-0629-3}

\bibitem[{{Cheung} {et~al.}(2022){Cheung}, {Mart{\'\i}nez-Sykora}, {Testa}, {De Pontieu}, {Chintzoglou}, {Rempel}, {Polito}, {Kerr}, {Reeves}, {Fletcher}, {Jin}, {N{\'o}brega-Siverio}, {Danilovic}, {Antolin}, {Allred}, {Hansteen}, {Ugarte-Urra}, {DeLuca}, {Longcope}, {Takasao}, {DeRosa}, {Boerner}, {Jaeggli}, {Nitta}, {Daw}, {Carlsson}, {Golub}, \& {The}}]{2022Cheung}
{Cheung}, M. C.~M., {Mart{\'\i}nez-Sykora}, J., {Testa}, P., {et~al.} 2022, \apj, 926, 53, \dodoi{10.3847/1538-4357/ac4223}

\bibitem[{{Culhane} {et~al.}(1981){Culhane}, {Rapley}, {Bentley}, {Gabriel}, {Phillips}, {Acton}, {Wolfson}, {Catura}, {Jordan}, \& {Antonucci}}]{1981Culhane}
{Culhane}, J.~L., {Rapley}, C.~G., {Bentley}, R.~D., {et~al.} 1981, \apjl, 244, L141, \dodoi{10.1086/183499}

\bibitem[{{Culhane} {et~al.}(1991){Culhane}, {Hiei}, {Doschek}, {Cruise}, {Ogawara}, {Uchida}, {Bentley}, {Brown}, {Lang}, {Watanabe}, {Bowles}, {Deslattes}, {Feldman}, {Fludra}, {Guttridge}, {Henins}, {Lapington}, {Magraw}, {Mariska}, {Payne}, {Phillips}, {Sheather}, {Slater}, {Tanaka}, {Towndrow}, {Trow}, \& {Yamaguchi}}]{1991Culhane}
{Culhane}, J.~L., {Hiei}, E., {Doschek}, G.~A., {et~al.} 1991, \solphys, 136, 89, \dodoi{10.1007/BF00151696}

\bibitem[{{Culhane} {et~al.}(2007){Culhane}, {Harra}, {James}, {Al-Janabi}, {Bradley}, {Chaudry}, {Rees}, {Tandy}, {Thomas}, {Whillock}, {Winter}, {Doschek}, {Korendyke}, {Brown}, {Myers}, {Mariska}, {Seely}, {Lang}, {Kent}, {Shaughnessy}, {Young}, {Simnett}, {Castelli}, {Mahmoud}, {Mapson-Menard}, {Probyn}, {Thomas}, {Davila}, {Dere}, {Windt}, {Shea}, {Hagood}, {Moye}, {Hara}, {Watanabe}, {Matsuzaki}, {Kosugi}, {Hansteen}, \& {Wikstol}}]{2007Culhane}
{Culhane}, J.~L., {Harra}, L.~K., {James}, A.~M., {et~al.} 2007, \solphys, 243, 19, \dodoi{10.1007/s01007-007-0293-1}

\bibitem[{{Dahlin} {et~al.}(2015){Dahlin}, {Drake}, \& {Swisdak}}]{2015Dahlin}
{Dahlin}, J.~T., {Drake}, J.~F., \& {Swisdak}, M. 2015, Physics of Plasmas, 22, 100704, \dodoi{10.1063/1.4933212}

\bibitem[{{Dahlin} {et~al.}(2017){Dahlin}, {Drake}, \& {Swisdak}}]{2017Dahlin}
---. 2017, Physics of Plasmas, 24, 092110, \dodoi{10.1063/1.4986211}

\bibitem[{{De Pontieu} {et~al.}(2020){De Pontieu}, {Mart{\'\i}nez-Sykora}, {Testa}, {Winebarger}, {Daw}, {Hansteen}, {Cheung}, \& {Antolin}}]{2020DePontieu}
{De Pontieu}, B., {Mart{\'\i}nez-Sykora}, J., {Testa}, P., {et~al.} 2020, \apj, 888, 3, \dodoi{10.3847/1538-4357/ab5b03}

\bibitem[{{De Pontieu} {et~al.}(2014){De Pontieu}, {Title}, {Lemen}, {Kushner}, {Akin}, {Allard}, {Berger}, {Boerner}, {Cheung}, {Chou}, {Drake}, {Duncan}, {Freeland}, {Heyman}, {Hoffman}, {Hurlburt}, {Lindgren}, {Mathur}, {Rehse}, {Sabolish}, {Seguin}, {Schrijver}, {Tarbell}, {W{\"u}lser}, {Wolfson}, {Yanari}, {Mudge}, {Nguyen-Phuc}, {Timmons}, {van Bezooijen}, {Weingrod}, {Brookner}, {Butcher}, {Dougherty}, {Eder}, {Knagenhjelm}, {Larsen}, {Mansir}, {Phan}, {Boyle}, {Cheimets}, {DeLuca}, {Golub}, {Gates}, {Hertz}, {McKillop}, {Park}, {Perry}, {Podgorski}, {Reeves}, {Saar}, {Testa}, {Tian}, {Weber}, {Dunn}, {Eccles}, {Jaeggli}, {Kankelborg}, {Mashburn}, {Pust}, {Springer}, {Carvalho}, {Kleint}, {Marmie}, {Mazmanian}, {Pereira}, {Sawyer}, {Strong}, {Worden}, {Carlsson}, {Hansteen}, {Leenaarts}, {Wiesmann}, {Aloise}, {Chu}, {Bush}, {Scherrer}, {Brekke}, {Martinez-Sykora}, {Lites}, {McIntosh}, {Uitenbroek}, {Okamoto}, {Gummin}, {Auker}, {Jerram}, {Pool}, \& {Waltham}}]{2014DePontieu}
{De Pontieu}, B., {Title}, A.~M., {Lemen}, J.~R., {et~al.} 2014, \solphys, 289, 2733, \dodoi{10.1007/s11207-014-0485-y}

\bibitem[{{De Pontieu} {et~al.}(2021){De Pontieu}, {Polito}, {Hansteen}, {Testa}, {Reeves}, {Antolin}, {N{\'o}brega-Siverio}, {Kowalski}, {Martinez-Sykora}, {Carlsson}, {McIntosh}, {Liu}, {Daw}, \& {Kankelborg}}]{2021DePontieu}
{De Pontieu}, B., {Polito}, V., {Hansteen}, V., {et~al.} 2021, \solphys, 296, 84, \dodoi{10.1007/s11207-021-01826-0}

\bibitem[{{Dolla} \& {Solomon}(2008)}]{2008DollaSolomon}
{Dolla}, L., \& {Solomon}, J. 2008, \aap, 483, 271, \dodoi{10.1051/0004-6361:20077903}

\bibitem[{{Doschek}(1990)}]{1990Doschek}
{Doschek}, G.~A. 1990, \apjs, 73, 117, \dodoi{10.1086/191443}

\bibitem[{{Doschek} {et~al.}(1981){Doschek}, {Feldman}, {Landecker}, \& {McKenzie}}]{1981Doschek}
{Doschek}, G.~A., {Feldman}, U., {Landecker}, P.~B., \& {McKenzie}, D.~L. 1981, \apj, 249, 372, \dodoi{10.1086/159294}

\bibitem[{{Doschek} {et~al.}(1979){Doschek}, {Kreplin}, \& {Feldman}}]{1979Doschek}
{Doschek}, G.~A., {Kreplin}, R.~W., \& {Feldman}, U. 1979, \apjl, 233, L157, \dodoi{10.1086/183096}

\bibitem[{{Doschek} {et~al.}(2014){Doschek}, {McKenzie}, \& {Warren}}]{2014Doschek}
{Doschek}, G.~A., {McKenzie}, D.~E., \& {Warren}, H.~P. 2014, \apj, 788, 26, \dodoi{10.1088/0004-637X/788/1/26}

\bibitem[{{Doschek} {et~al.}(1986){Doschek}, {Antiochos}, {Antonucci}, {Cheng}, {Culhane}, {Fisher}, {Jordan}, {Leibacher}, {MacNiece}, {McWhirter}, {Moore}, {Rabin}, {Rust}, \& {Shine}}]{1986Doschek}
{Doschek}, G.~A., {Antiochos}, S.~K., {Antonucci}, E., {et~al.} 1986, in NASA Conference Publication, Vol. 2439, NASA Conference Publication, 4

\bibitem[{{Drake} {et~al.}(2009{\natexlab{a}}){Drake}, {Cassak}, {Shay}, {Swisdak}, \& {Quataert}}]{2009DrakeApJ}
{Drake}, J.~F., {Cassak}, P.~A., {Shay}, M.~A., {Swisdak}, M., \& {Quataert}, E. 2009{\natexlab{a}}, \apjl, 700, L16, \dodoi{10.1088/0004-637X/700/1/L16}

\bibitem[{{Drake} \& {Swisdak}(2014)}]{2014DrakeSwisdak}
{Drake}, J.~F., \& {Swisdak}, M. 2014, Physics of Plasmas, 21, 072903, \dodoi{10.1063/1.4889871}

\bibitem[{{Drake} {et~al.}(2009{\natexlab{b}}){Drake}, {Swisdak}, {Phan}, {Cassak}, {Shay}, {Lepri}, {Lin}, {Quataert}, \& {Zurbuchen}}]{2009Drake}
{Drake}, J.~F., {Swisdak}, M., {Phan}, T.~D., {et~al.} 2009{\natexlab{b}}, Journal of Geophysical Research (Space Physics), 114, A05111, \dodoi{10.1029/2008JA013701}

\bibitem[{{Feldman} {et~al.}(2005){Feldman}, {Landi}, \& {Laming}}]{2005Feldman}
{Feldman}, U., {Landi}, E., \& {Laming}, J.~M. 2005, \apj, 619, 1142, \dodoi{10.1086/426539}

\bibitem[{{Fletcher}(2024)}]{2024Fletcher}
{Fletcher}, L. 2024, \araa, 62, 437, \dodoi{10.1146/annurev-astro-052920-010547}

\bibitem[{{Fletcher} {et~al.}(2011){Fletcher}, {Dennis}, {Hudson}, {Krucker}, {Phillips}, {Veronig}, {Battaglia}, {Bone}, {Caspi}, {Chen}, {Gallagher}, {Grigis}, {Ji}, {Liu}, {Milligan}, \& {Temmer}}]{2011Fletcher}
{Fletcher}, L., {Dennis}, B.~R., {Hudson}, H.~S., {et~al.} 2011, \ssr, 159, 19, \dodoi{10.1007/s11214-010-9701-8}

\bibitem[{{Fludra} \& {Schmelz}(1999)}]{1999FludraSchmelz}
{Fludra}, A., \& {Schmelz}, J.~T. 1999, \aap, 348, 286

\bibitem[{{French} {et~al.}(2024){French}, {Yu}, {Chen}, {Shen}, \& {Matthews}}]{2024French}
{French}, R.~J., {Yu}, S., {Chen}, B., {Shen}, C., \& {Matthews}, S.~A. 2024, \mnras, 528, 6836, \dodoi{10.1093/mnras/stae430}

\bibitem[{{Haggerty} {et~al.}(2015){Haggerty}, {Shay}, {Drake}, {Phan}, \& {McHugh}}]{2015Haggerty}
{Haggerty}, C.~C., {Shay}, M.~A., {Drake}, J.~F., {Phan}, T.~D., \& {McHugh}, C.~T. 2015, \grl, 42, 9657, \dodoi{10.1002/2015GL065961}

\bibitem[{{Hara} {et~al.}(2008){Hara}, {Watanabe}, {Matsuzaki}, {Harra}, {Culhane}, {Cargill}, {Mariska}, \& {Doschek}}]{2008Hara}
{Hara}, H., {Watanabe}, T., {Matsuzaki}, K., {et~al.} 2008, \pasj, 60, 275, \dodoi{10.1093/pasj/60.2.275}

\bibitem[{{Harra} {et~al.}(2013){Harra}, {Matthews}, {Culhane}, {Cheung}, {Kontar}, \& {Hara}}]{2013Harra}
{Harra}, L.~K., {Matthews}, S., {Culhane}, J.~L., {et~al.} 2013, \apj, 774, 122, \dodoi{10.1088/0004-637X/774/2/122}

\bibitem[{{Hernandez} \& {Marsch}(1985)}]{1985HernandezMarsch}
{Hernandez}, R., \& {Marsch}, E. 1985, \jgr, 90, 11062, \dodoi{10.1029/JA090iA11p11062}

\bibitem[{{Hudson} {et~al.}(2021){Hudson}, {Sim{\~o}es}, {Fletcher}, {Hayes}, \& {Hannah}}]{2021Hudson}
{Hudson}, H.~S., {Sim{\~o}es}, P. J.~A., {Fletcher}, L., {Hayes}, L.~A., \& {Hannah}, I.~G. 2021, \mnras, 501, 1273, \dodoi{10.1093/mnras/staa3664}

\bibitem[{{Imada} {et~al.}(2009){Imada}, {Hara}, \& {Watanabe}}]{2009Imada}
{Imada}, S., {Hara}, H., \& {Watanabe}, T. 2009, \apjl, 705, L208, \dodoi{10.1088/0004-637X/705/2/L208}

\bibitem[{{Knizhnik} {et~al.}(2011){Knizhnik}, {Swisdak}, \& {Drake}}]{2011Knizhnik}
{Knizhnik}, K., {Swisdak}, M., \& {Drake}, J.~F. 2011, \apjl, 743, L35, \dodoi{10.1088/2041-8205/743/2/L35}

\bibitem[{{Kontar} {et~al.}(2017){Kontar}, {Perez}, {Harra}, {Kuznetsov}, {Emslie}, {Jeffrey}, {Bian}, \& {Dennis}}]{2017Kontar}
{Kontar}, E.~P., {Perez}, J.~E., {Harra}, L.~K., {et~al.} 2017, \prl, 118, 155101, \dodoi{10.1103/PhysRevLett.118.155101}

\bibitem[{{Landi}(2007)}]{2007Landi}
{Landi}, E. 2007, \apj, 663, 1363, \dodoi{10.1086/517910}

\bibitem[{{Larosa} \& {Moore}(1993)}]{1993Larosa}
{Larosa}, T.~N., \& {Moore}, R.~L. 1993, \apj, 418, 912, \dodoi{10.1086/173448}

\bibitem[{{Larosa} {et~al.}(1994){Larosa}, {Moore}, \& {Shore}}]{1994Larosa}
{Larosa}, T.~N., {Moore}, R.~L., \& {Shore}, S.~N. 1994, \apj, 425, 856, \dodoi{10.1086/174031}

\bibitem[{{Mariska}(1994)}]{1994Mariska}
{Mariska}, J.~T. 1994, \apj, 434, 756, \dodoi{10.1086/174778}

\bibitem[{{Mariska} {et~al.}(1993){Mariska}, {Doschek}, \& {Bentley}}]{1993Mariska}
{Mariska}, J.~T., {Doschek}, G.~A., \& {Bentley}, R.~D. 1993, \apj, 419, 418, \dodoi{10.1086/173494}

\bibitem[{{McKenzie} {et~al.}(1980){McKenzie}, {Broussard}, {Landecker}, {Rugge}, {Young}, {Doschek}, \& {Feldman}}]{1980McKenzie}
{McKenzie}, D.~L., {Broussard}, R.~M., {Landecker}, P.~B., {et~al.} 1980, \apjl, 238, L43, \dodoi{10.1086/183254}

\bibitem[{{Mistry} {et~al.}(2017){Mistry}, {Eastwood}, {Phan}, \& {Hietala}}]{2017Mistry}
{Mistry}, R., {Eastwood}, J.~P., {Phan}, T.~D., \& {Hietala}, H. 2017, Journal of Geophysical Research (Space Physics), 122, 5895, \dodoi{10.1002/2017JA024032}

\bibitem[{{{\O}ieroset} {et~al.}(2023){{\O}ieroset}, {Phan}, {Oka}, {Drake}, {Fuselier}, {Gershman}, {Maheshwari}, {Giles}, {Zhang}, {Guo}, {Burch}, {Torbert}, \& {Strangeway}}]{2023Oieroset}
{{\O}ieroset}, M., {Phan}, T.~D., {Oka}, M., {et~al.} 2023, \apj, 954, 118, \dodoi{10.3847/1538-4357/acdf44}

\bibitem[{{{\O}ieroset} {et~al.}(2024){{\O}ieroset}, {Phan}, {Drake}, {Starkey}, {Fuselier}, {Cohen}, {Haggerty}, {Shay}, {Oka}, {Gershman}, {Maheshwari}, {Burch}, {Torbert}, \& {Strangeway}}]{2024Oieroset}
{{\O}ieroset}, M., {Phan}, T.~D., {Drake}, J.~F., {et~al.} 2024, \apj, 971, 144, \dodoi{10.3847/1538-4357/ad6151}

\bibitem[{{Oka} {et~al.}(2025){Oka}, {Phan}, {{\O}ieroset}, {Gershman}, {Torbert}, {Burch}, \& {Angelopoulos}}]{2025Oka}
{Oka}, M., {Phan}, T.~D., {{\O}ieroset}, M., {et~al.} 2025, \apj, 984, 150, \dodoi{10.3847/1538-4357/adc5e5}

\bibitem[{{Phan} {et~al.}(2013){Phan}, {Shay}, {Gosling}, {Fujimoto}, {Drake}, {Paschmann}, {Oieroset}, {Eastwood}, \& {Angelopoulos}}]{2013Phan}
{Phan}, T.~D., {Shay}, M.~A., {Gosling}, J.~T., {et~al.} 2013, \grl, 40, 4475, \dodoi{10.1002/grl.50917}

\bibitem[{{Phan} {et~al.}(2014){Phan}, {Drake}, {Shay}, {Gosling}, {Paschmann}, {Eastwood}, {Oieroset}, {Fujimoto}, \& {Angelopoulos}}]{2014Phan}
{Phan}, T.~D., {Drake}, J.~F., {Shay}, M.~A., {et~al.} 2014, \grl, 41, 7002, \dodoi{10.1002/2014GL061547}

\bibitem[{{Phillips} \& {Dennis}(2012)}]{2012PhillipsDennis}
{Phillips}, K.~J.~H., \& {Dennis}, B.~R. 2012, \apj, 748, 52, \dodoi{10.1088/0004-637X/748/1/52}

\bibitem[{{Polito} {et~al.}(2019){Polito}, {Testa}, \& {De Pontieu}}]{2019Polito}
{Polito}, V., {Testa}, P., \& {De Pontieu}, B. 2019, \apjl, 879, L17, \dodoi{10.3847/2041-8213/ab290b}

\bibitem[{{Ruan} {et~al.}(2023){Ruan}, {Yan}, \& {Keppens}}]{2023Ruan}
{Ruan}, W., {Yan}, L., \& {Keppens}, R. 2023, \apj, 947, 67, \dodoi{10.3847/1538-4357/ac9b4e}

\bibitem[{{Russell}(2024)}]{2024Russell}
{Russell}, A. J.~B. 2024, Geophysical Monograph Series, 285, 39, \dodoi{10.1002/9781394195985.ch3}

\bibitem[{{Seely} {et~al.}(1997){Seely}, {Feldman}, {Sch{\"u}hle}, {Wilhelm}, {Curdt}, \& {Lemaire}}]{1997Seely}
{Seely}, J.~F., {Feldman}, U., {Sch{\"u}hle}, U., {et~al.} 1997, \apjl, 484, L87, \dodoi{10.1086/310769}

\bibitem[{{Shay} {et~al.}(2014){Shay}, {Haggerty}, {Phan}, {Drake}, {Cassak}, {Wu}, {Oieroset}, {Swisdak}, \& {Malakit}}]{2014Shay}
{Shay}, M.~A., {Haggerty}, C.~C., {Phan}, T.~D., {et~al.} 2014, Physics of Plasmas, 21, 122902, \dodoi{10.1063/1.4904203}

\bibitem[{{Shen} {et~al.}(2023){Shen}, {Polito}, {Reeves}, {Chen}, {Yu}, \& {Xie}}]{2023Shen}
{Shen}, C., {Polito}, V., {Reeves}, K.~K., {et~al.} 2023, Frontiers in Astronomy and Space Sciences, 10, 19, \dodoi{10.3389/fspas.2023.1096133}

\bibitem[{{Shibata} {et~al.}(2023){Shibata}, {Takasao}, \& {Reeves}}]{2023Shibata}
{Shibata}, K., {Takasao}, S., \& {Reeves}, K.~K. 2023, \apj, 943, 106, \dodoi{10.3847/1538-4357/acaa9c}

\bibitem[{{Shimizu} {et~al.}(2019){Shimizu}, {Imada}, {Kawate}, {Ichimoto}, {Suematsu}, {Hara}, {Katsukawa}, {Kubo}, {Toriumi}, {Watanabe}, {Yokoyama}, {Korendyke}, {Warren}, {Tarbell}, {De Pontieu}, {Teriaca}, {Sch{\"u}hle}, {Solanki}, {Harra}, {Matthews}, {Fludra}, {Auch{\`e}re}, {Andretta}, {Naletto}, \& {Zhukov}}]{2019Shimizu}
{Shimizu}, T., {Imada}, S., {Kawate}, T., {et~al.} 2019, in Society of Photo-Optical Instrumentation Engineers (SPIE) Conference Series, Vol. 11118, UV, X-Ray, and Gamma-Ray Space Instrumentation for Astronomy XXI, ed. O.~H. {Siegmund}, 1111807, \dodoi{10.1117/12.2528240}

\bibitem[{{Stores} {et~al.}(2021){Stores}, {Jeffrey}, \& {Kontar}}]{2021Stores}
{Stores}, M., {Jeffrey}, N. L.~S., \& {Kontar}, E.~P. 2021, \apj, 923, 40, \dodoi{10.3847/1538-4357/ac2c65}

\bibitem[{{Sylwester} {et~al.}(2022){Sylwester}, {Sylwester}, {Phillips}, \& {Kepa}}]{2022Sylwester}
{Sylwester}, J., {Sylwester}, B., {Phillips}, K.~J.~H., \& {Kepa}, A. 2022, \apj, 930, 77, \dodoi{10.3847/1538-4357/ac5b0d}

\bibitem[{{Tanaka}(1986)}]{1986Tanaka}
{Tanaka}, K. 1986, \pasj, 38, 225

\bibitem[{{Tanaka} {et~al.}(1982){Tanaka}, {Watanabe}, {Nishi}, \& {Akita}}]{1982Tanaka}
{Tanaka}, K., {Watanabe}, T., {Nishi}, K., \& {Akita}, K. 1982, \apjl, 254, L59, \dodoi{10.1086/183756}

\bibitem[{{Tilquin} {et~al.}(2020){Tilquin}, {Eastwood}, \& {Phan}}]{2020Tilquin}
{Tilquin}, H., {Eastwood}, J.~P., \& {Phan}, T.~D. 2020, \apj, 895, 68, \dodoi{10.3847/1538-4357/ab8812}

\bibitem[{{Tracy} {et~al.}(2015){Tracy}, {Kasper}, {Zurbuchen}, {Raines}, {Shearer}, \& {Gilbert}}]{2015Tracy}
{Tracy}, P.~J., {Kasper}, J.~C., {Zurbuchen}, T.~H., {et~al.} 2015, \apj, 812, 170, \dodoi{10.1088/0004-637X/812/2/170}

\bibitem[{{Tsuneta}(1995)}]{1995Tsuneta}
{Tsuneta}, S. 1995, \pasj, 47, 691

\bibitem[{{Tu} {et~al.}(1998){Tu}, {Marsch}, {Wilhelm}, \& {Curdt}}]{1998Tu}
{Tu}, C.~Y., {Marsch}, E., {Wilhelm}, K., \& {Curdt}, W. 1998, \apj, 503, 475, \dodoi{10.1086/305982}

\bibitem[{{van der Holst} {et~al.}(2014){van der Holst}, {Sokolov}, {Meng}, {Jin}, {Manchester}, {T{\'o}th}, \& {Gombosi}}]{2014vanderHolst}
{van der Holst}, B., {Sokolov}, I.~V., {Meng}, X., {et~al.} 2014, \apj, 782, 81, \dodoi{10.1088/0004-637X/782/2/81}

\bibitem[{{van der Holst} {et~al.}(2022){van der Holst}, {Huang}, {Sachdeva}, {Kasper}, {Manchester}, {Borovikov}, {Chandran}, {Case}, {Korreck}, {Larson}, {Livi}, {Stevens}, {Whittlesey}, {Bale}, {Pulupa}, {Malaspina}, {Bonnell}, {Harvey}, {Goetz}, \& {MacDowall}}]{2022vanderHolst}
{van der Holst}, B., {Huang}, J., {Sachdeva}, N., {et~al.} 2022, \apj, 925, 146, \dodoi{10.3847/1538-4357/ac3d34}

\bibitem[{{Yin} {et~al.}(2024){Yin}, {Drake}, \& {Swisdak}}]{2024Yin}
{Yin}, Z., {Drake}, J.~F., \& {Swisdak}, M. 2024, \apj, 974, 74, \dodoi{10.3847/1538-4357/ad7131}

\bibitem[{{Zhang} {et~al.}(2019){Zhang}, {Drake}, \& {Swisdak}}]{2019Zhang}
{Zhang}, Q., {Drake}, J.~F., \& {Swisdak}, M. 2019, Physics of Plasmas, 26, 072115, \dodoi{10.1063/1.5104352}

\bibitem[{{Zhang} {et~al.}(2021){Zhang}, {Guo}, {Daughton}, {Li}, \& {Li}}]{2021Zhang}
{Zhang}, Q., {Guo}, F., {Daughton}, W., {Li}, H., \& {Li}, X. 2021, \prl, 127, 185101, \dodoi{10.1103/PhysRevLett.127.185101}

\bibitem[{{Zhu} {et~al.}(2023){Zhu}, {Szente}, \& {Landi}}]{2023Zhu}
{Zhu}, Y., {Szente}, J., \& {Landi}, E. 2023, \apj, 948, 90, \dodoi{10.3847/1538-4357/acc187}

\end{thebibliography}
\bibliographystyle{aasjournal}

\end{document}